\documentclass{svmult}
\usepackage{epsfig,graphicx} 
\usepackage[bottom]{footmisc}
\usepackage{natbib,url}      
\bibpunct{(}{)}{;}{a}{}{,}   
\def\cite#1{\citealp{#1}}    

\begin{document}


\def\thisvolume{these proceedings}

\def\aj{{AJ}}			
\def\araa{{ARA\&A}}		
\def\apj{{ApJ}}			
\def\apjl{{ApJ}}		
\def\apjs{{ApJS}}		
\def\ao{{Appl.\ Optics}} 
\def\apss{{Ap\&SS}}		
\def\aap{{A\&A}}		
\def\aapr{{A\&A~Rev.}}		
\def\aaps{{A\&AS}}		
\def\an{{Astron.\ Nachrichten}}
\def\aspcs{{ASP Conf.\ Ser.}}
\def\azh{{AZh}}			
\def\baas{{BAAS}}		
\def\jrasc{{JRASC}}		
\def\memras{{MmRAS}}		
\def\mnras{{MNRAS}}
\def\nat{{Nat}}		
\def\pra{{Phys.\ Rev.\ A}} 
\def\prb{{Phys.\ Rev.\ B}}		
\def\prc{{Phys.\ Rev.\ C}}		
\def\prd{{Phys.\ Rev.\ D}}		
\def\prl{{Phys.\ Rev.\ Lett}}	
\def\pasp{{PASP}}
\def\pasj{{PASJ}}		
\def\qjras{{QJRAS}}
\def\science{{Sci}}		
\def\skytel{{S\&T}}		
\def\solphys{{Solar\ Phys.}} 
\def\sovast{{Soviet\ Ast.}}  
\def\ssr{{Space\ Sci.\ Rev.}}
\def\svassp{{Astrophys.\ Space Science Proc.}}
\def\zap{{ZAp}}			
\let\astap=\aap
\let\apjlett=\apjl
\let\apjsupp=\apjs

\def\ion#1#2{{\rm #1}\,{\uppercase{#2}}}  
\def\deg{\hbox{$^\circ$}}
\def\sun{\hbox{$\odot$}}
\def\earth{\hbox{$\oplus$}}
\def\la{\mathrel{\hbox{\rlap{\hbox{\lower4pt\hbox{$\sim$}}}\hbox{$<$}}}}
\def\ga{\mathrel{\hbox{\rlap{\hbox{\lower4pt\hbox{$\sim$}}}\hbox{$>$}}}}
\def\sq{\hbox{\rlap{$\sqcap$}$\sqcup$}}
\def\arcmin{\hbox{$^\prime$}}
\def\arcsec{\hbox{$^{\prime\prime}$}}
\def\fd{\hbox{$.\!\!^{\rm d}$}}
\def\fh{\hbox{$.\!\!^{\rm h}$}}
\def\fm{\hbox{$.\!\!^{\rm m}$}}
\def\fs{\hbox{$.\!\!^{\rm s}$}}
\def\fdg{\hbox{$.\!\!^\circ$}}
\def\farcm{\hbox{$.\mkern-4mu^\prime$}}
\def\farcs{\hbox{$.\!\!^{\prime\prime}$}}
\def\fp{\hbox{$.\!\!^{\scriptscriptstyle\rm p}$}}
\def\micron{\hbox{$\mu$m}}
\def\onehalf{\hbox{$\,^1\!/_2$}}	
\def\onethird{\hbox{$\,^1\!/_3$}}
\def\twothirds{\hbox{$\,^2\!/_3$}}
\def\onequarter{\hbox{$\,^1\!/_4$}}
\def\threequarters{\hbox{$\,^3\!/_4$}}
\def\ubv{\hbox{$U\!BV$}}		
\def\ubvr{\hbox{$U\!BV\!R$}}		
\def\ubvri{\hbox{$U\!BV\!RI$}}		
\def\ubvrij{\hbox{$U\!BV\!RI\!J$}}		
\def\ubvrijh{\hbox{$U\!BV\!RI\!J\!H$}}		
\def\ubvrijhk{\hbox{$U\!BV\!RI\!J\!H\!K$}}		
\def\ub{\hbox{$U\!-\!B$}}		
\def\bv{\hbox{$B\!-\!V$}}		
\def\vr{\hbox{$V\!-\!R$}}		
\def\ur{\hbox{$U\!-\!R$}}


\def\labelitemi{{\bf --}}  

\def\rmit#1{{\it #1}}              
\def\rmit#1{{\rm #1}}              
\def\etal{\rmit{et al.}}           
\def\etc{\rmit{etc.}}           
\def\ie{\rmit{i.e.,}}              
\def\eg{\rmit{e.g.,}}              
\def\cf{cf.}                       
\def\viz{\rmit{viz.}}
\def\vs{\rmit{vs.}}

\def\rot{\hbox{\rm rot}}
\def\div{\hbox{\rm div}}
\def\lesssim{\mathrel{\hbox{\rlap{\hbox{\lower4pt\hbox{$\sim$}}}\hbox{$<$}}}}
\def\gtrsim{\mathrel{\hbox{\rlap{\hbox{\lower4pt\hbox{$\sim$}}}\hbox{$>$}}}}
\def\dif{\: {\rm d}}                       
\def\ep{\:{\rm e}^}                        
\def\dash{\hbox{$\,-\,$}}                  
\def\is{\!=\!}                             

\def\starname#1#2{${#1}$\,{\rm {#2}}}  
\def\Teff{\hbox{$T_{\rm eff}$}}   

\def\kms{\hbox{km$\;$s$^{-1}$}}
\def\Mxcm{\hbox{Mx\,cm$^{-2}$}}    

\def\Bapp{\hbox{$B_{\rm app}$}}    

\def\komega{($k, \omega$)}                 
\def\kf{($k_h,f$)}                         
\def\VminI{\hbox{$V\!\!-\!\!I$}}           
\def\IminI{\hbox{$I\!\!-\!\!I$}}           
\def\VminV{\hbox{$V\!\!-\!\!V$}}           
\def\Xt{\hbox{$X\!\!-\!t$}}                

\def\level #1 #2#3#4{$#1 \: ^{#2} \mbox{#3} ^{#4}$}   

\def\specchar#1{\uppercase{#1}}    
\def\AlI{\mbox{Al\,\specchar{i}}}  
\def\BI{\mbox{B\,\specchar{i}}} 
\def\BII{\mbox{B\,\specchar{ii}}}  
\def\BaI{\mbox{Ba\,\specchar{i}}}  
\def\BaII{\mbox{Ba\,\specchar{ii}}} 
\def\CI{\mbox{C\,\specchar{i}}} 
\def\CII{\mbox{C\,\specchar{ii}}} 
\def\CIII{\mbox{C\,\specchar{iii}}} 
\def\CIV{\mbox{C\,\specchar{iv}}} 
\def\CaI{\mbox{Ca\,\specchar{i}}} 
\def\CaII{\mbox{Ca\,\specchar{ii}}} 
\def\CaIII{\mbox{Ca\,\specchar{iii}}} 
\def\CoI{\mbox{Co\,\specchar{i}}} 
\def\CrI{\mbox{Cr\,\specchar{i}}} 
\def\CriI{\mbox{Cr\,\specchar{ii}}} 
\def\CsI{\mbox{Cs\,\specchar{i}}} 
\def\CsII{\mbox{Cs\,\specchar{ii}}} 
\def\CuI{\mbox{Cu\,\specchar{i}}} 
\def\FeI{\mbox{Fe\,\specchar{i}}} 
\def\FeII{\mbox{Fe\,\specchar{ii}}} 
\def\FeIX{\mbox{Fe\,\specchar{ix}}}
\def\FeX{\mbox{Fe\,\specchar{x}}}
\def\FeXVI{\mbox{Fe\,\specchar{xvi}}}
\def\FrI{\mbox{Fr\,\specchar{i}}}
\def\HI{\mbox{H\,\specchar{i}}} 
\def\HII{\mbox{H\,\specchar{ii}}} 
\def\Hmin{\hbox{\rmH$^{^{_{\scriptstyle -}}}$}}      
\def\Hemin{\hbox{{\rm He}$^{^{_{\scriptstyle -}}}$}} 
\def\HeI{\mbox{He\,\specchar{i}}} 
\def\HeII{\mbox{He\,\specchar{ii}}} 
\def\HeIII{\mbox{He\,\specchar{iii}}} 
\def\KI{\mbox{K\,\specchar{i}}} 
\def\KII{\mbox{K\,\specchar{ii}}} 
\def\KIII{\mbox{K\,\specchar{iii}}} 
\def\LiI{\mbox{Li\,\specchar{i}}} 
\def\LiII{\mbox{Li\,\specchar{ii}}} 
\def\LiIII{\mbox{Li\,\specchar{iii}}} 
\def\MgI{\mbox{Mg\,\specchar{i}}} 
\def\MgII{\mbox{Mg\,\specchar{ii}}} 
\def\MgIII{\mbox{Mg\,\specchar{iii}}} 
\def\MnI{\mbox{Mn\,\specchar{i}}} 
\def\NI{\mbox{N\,\specchar{i}}}
\def\NaI{\mbox{Na\,\specchar{i}}}
\def\NaII{\mbox{Na\,\specchar{ii}}}
\def\NaIII{\mbox{Na\,\specchar{iii}}} 
\def\NiI{\mbox{Ni\,\specchar{i}}} 
\def\NiII{\mbox{Ni\,\specchar{ii}}}
\def\NiIII{\mbox{Ni\,\specchar{iii}}} 
\def\OI{\mbox{O\,\specchar{i}}} 
\def\OVI{\mbox{O\,\specchar{vi}}}
\def\RbI{\mbox{Rb\,\specchar{i}}} 
\def\SII{\mbox{S\,\specchar{ii}}} 
\def\SiI{\mbox{Si\,\specchar{i}}} 
\def\SiII{\mbox{Si\,\specchar{ii}}} 
\def\SrI{\mbox{Sr\,\specchar{i}}}
\def\SrII{\mbox{Sr\,\specchar{ii}}}
\def\TiI{\mbox{Ti\,\specchar{i}}} 
\def\TiII{\mbox{Ti\,\specchar{ii}}} 
\def\TiIII{\mbox{Ti\,\specchar{iii}}} 
\def\TiIV{\mbox{Ti\,\specchar{iv}}} 
\def\VI{\mbox{V\,\specchar{i}}} 
\def\HtwoO{\mbox{H$_2$O}}        
\def\Otwo{\mbox{O$_2$}}          

\def\Halpha{\mbox{H\hspace{0.1ex}$\alpha$}} 
\def\Ha{\mbox{H\hspace{0.2ex}$\alpha$}}
\def\Hbeta{\mbox{H\hspace{0.2ex}$\beta$}}
\def\Hgamma{\mbox{H\hspace{0.2ex}$\gamma$}}
\def\Hdelta{\mbox{H\hspace{0.2ex}$\delta$}}
\def\Hepsilon{\mbox{H\hspace{0.2ex}$\epsilon$}}
\def\Hzeta{\mbox{H\hspace{0.2ex}$\zeta$}}
\def\Lyalpha{\mbox{Ly$\hspace{0.2ex}\alpha$}}
\def\Lybeta{\mbox{Ly$\hspace{0.2ex}\beta$}}
\def\Lygamma{\mbox{Ly$\hspace{0.2ex}\gamma$}}
\def\Lycont{\mbox{Ly\hspace{0.2ex}{\small cont}}}
\def\Baalpha{\mbox{Ba$\hspace{0.2ex}\alpha$}}
\def\Babeta{\mbox{Ba$\hspace{0.2ex}\beta$}}
\def\Bacont{\mbox{Ba\hspace{0.2ex}{\small cont}}}
\def\Paalpha{\mbox{Pa$\hspace{0.2ex}\alpha$}}
\def\Bralpha{\mbox{Br$\hspace{0.2ex}\alpha$}}

\def\NaD{\mbox{Na\,\specchar{i}\,D}}    
\def\NaDone{\mbox{Na\,\specchar{i}\,\,D$_1$}}
\def\NaDtwo{\mbox{Na\,\specchar{i}\,\,D$_2$}}
\def\NaID{\mbox{Na\,\specchar{i}\,\,D}}
\def\NaIDone{\mbox{Na\,\specchar{i}\,\,D$_1$}}
\def\NaIDtwo{\mbox{Na\,\specchar{i}\,\,D$_2$}}
\def\Done{\mbox{D$_1$}}
\def\Dtwo{\mbox{D$_2$}}

\def\Mgbone{\mbox{Mg\,\specchar{i}\,b$_1$}}
\def\Mgbtwo{\mbox{Mg\,\specchar{i}\,b$_2$}}
\def\Mgbthree{\mbox{Mg\,\specchar{i}\,b$_3$}}
\def\MgIb{\mbox{Mg\,\specchar{i}\,b}}
\def\MgIbone{\mbox{Mg\,\specchar{i}\,b$_1$}}
\def\MgIbtwo{\mbox{Mg\,\specchar{i}\,b$_2$}}
\def\MgIbthree{\mbox{Mg\,\specchar{i}\,b$_3$}}

\def\CaIIK{\mbox{Ca\,\specchar{ii}\,K}}       
\def\CaIIH{\mbox{Ca\,\specchar{ii}\,H}}
\def\CaIIHK{\mbox{Ca\,\specchar{ii}\,H\,\&\,K}}
\def\HK{\mbox{H\,\&\,K}}
\def\Kthree{\mbox{K$_3$}}      
\def\Hthree{\mbox{H$_3$}}
\def\Ktwo{\mbox{K$_2$}}
\def\Htwo{\mbox{H$_2$}}
\def\Kone{\mbox{K$_1$}}     
\def\Hone{\mbox{H$_1$}}     
\def\KtwoV{\mbox{K$_{2V}$}}
\def\KtwoR{\mbox{K$_{2R}$}}
\def\KoneV{\mbox{K$_{1V}$}}
\def\KoneR{\mbox{K$_{1R}$}}
\def\HtwoV{\mbox{H$_{2V}$}}
\def\HtwoR{\mbox{H$_{2R}$}}
\def\HoneV{\mbox{H$_{1V}$}}
\def\HoneR{\mbox{H$_{1R}$}}

\def\hk{\mbox{h\,\&\,k}}
\def\kthree{\mbox{k$_3$}}    
\def\hthree{\mbox{h$_3$}}
\def\ktwo{\mbox{k$_2$}}
\def\htwo{\mbox{h$_2$}}
\def\kone{\mbox{k$_1$}}     
\def\hone{\mbox{h$_1$}}     
\def\ktwoV{\mbox{k$_{2V}$}}
\def\ktwoR{\mbox{k$_{2R}$}}
\def\koneV{\mbox{k$_{1V}$}}
\def\koneR{\mbox{k$_{1R}$}}
\def\htwoV{\mbox{h$_{2V}$}}
\def\htwoR{\mbox{h$_{2R}$}}
\def\honeV{\mbox{h$_{1V}$}}
\def\honeR{\mbox{h$_{1R}$}}


\title*{Phase III of USO Solar Vector Magnetograph }


\author{Sanjay Gosain\inst{}
        \and
        P. Venkatakrishnan\inst{}}


\institute{Udaipur Solar Observatory, P. Box 198, Dewali, Udaipur-313001, Rajasthan, India.}

\maketitle


\begin{abstract}
  The Solar Vector Magnetograph (SVM) is a modern imaging spectropolarimeter installed at Udaipur Solar Observatory (USO). Earlier phases saw the development of the instrument using off-the-shelf components with in-house software development. Subsequently, improvements were done in the opto-mechanical design of the sub-systems and the telescope tracking system. The third phase of the instrument development saw three major improvements, these include: (i) installation of a web-camera based telescope guiding system, developed in-house, (ii) high-cadence spectropolarimetry using Liquid Crystal Variable Retarders and a fast CCD camera and (iii)inclusion of Na D1 line with regular photospheric Fe 630.2 nm line for chromospheric observations.
\end{abstract}

\section{Introduction}
\label{sec:introduction}
The Solar Vector Magnetograph is a filter-based imaging Stokes polarimeter. It takes monochromatic pictures of the solar active regions using a tunable narrow pass-band Fabry-Perot (FP) filter with a full width at half maximum (FWHM) of  about $\Delta\lambda = 120$~m\AA\ at $\lambda=6300$~\AA. The Stokes spectra over the field-of-view are constructed by scanning the FP filter. Polarization modulation was done with quarter waveplates during phase I \& II and in phase III Liquid Crystal Variable Retarders (LCVRs) are planned. The details of the instrument design parameters, hardware and software subsystems can be found in  \citet{2004ExA....18...31G,2006JApA...27..285G,2008JApA...29..107G}.  The main features of the instrument are as follows: (i) straight and symmetric optical design, (ii) direct pointing towards the Sun with no oblique reflections in the entire optical path and therefore minimal instrumental polarization, and (iii) dual-beam polarization analyzer for minimal seeing effects.

\section{Recent Developments}
\label{sec:recent}
{\bf (a) Webcam based guider system:} A low-cost webcam based guider system is developed for SVM. The telescope mount is German-Equatorial system equipped with MKS-4000 Telescope Control System. The guider-scope makes full-disk solar image on the webcam. An interactive software is developed in-house to grab images from the webcam and compute RA and DEC drifts due to telescope tracking errors. The errors are corrected by giving TTL pulse to the telescope drive. The residual rms guiding error is about 1 arc-sec).

\noindent{\bf (b) Comparison with HINODE SOT/SP Observations}
Comparison of spectro-polarimetric observations of a common location in a sunspot penumbral region was carried out between HINODE SOT/SP and SVM. In spite of polarization signals in SVM observations being diluted due to the combined effects of spatial and spectral resolution, leakage from adjacent Fabry-Perot transmission channels and smearing due to seeing, Milne-Eddington inversions of the two data-sets carried out using MELANIE inversion code (\cite{2001ASPC..236..487S})  give similar values for magnetic parameters (Table~\ref{table1}). These initial results indicate that the full Stokes inversion of SVM observations can yield reliable magnetic fields parameters. A more detailed comparison of the two instruments for a larger field-of-view is being carried out and will be reported in a forthcoming paper.

\begin{table}
\caption{Comparison of the physical parameters retrieved by inversion}
\centering
\begin{tabular}{lcccccr}
\hline
\newline
Parameter&~~~& ~~~~& SVM Udaipur  &~~~~ & ~~~~&HINODE SOT/SP  \\
\hline
\newline
Field Strength  (Gauss)~~~&~~ & ~~~~&  1302  $\pm$ 200& ~~~~& ~~~~&1277  $\pm$ 120  \\
Inclination ($^\circ$) &~~ & ~~~~& 	126   $\pm$ 12	&~~~~ &~~~~ &120   $\pm$ 8  \\
Stray Light&~~ & ~~~~&	0.38  $\pm$ 0.13	&~~~~ & ~~~~ &0.0   $\pm$ 0.0 \\
\hline
\newline
\label{table1}
\end{tabular}
\end{table}

\section{Ongoing Developments:}
\label{sec:ongoing}
{\bf (a) LCVR based polarization modulator:} LCVR modulators are procured from M/s Meadowlark Optics and integrated in the SVM optical scheme. The calibration of retardation voltages for different wavelengths and temperatures is under progress.

\noindent{\bf (b) Automated Polarimeter Calibration Unit:} The calibration of quarter waveplate based polarimeter during phase I \& II was done manually following POLIS scheme (\cite{2005A&A...437.1159B}). Currently this is being automated for regular calibration.

\noindent{\bf (c) Fast CCD camera \& Na D1 line:} Earlier camera Apogee6E had slow frame rate of 1 per second. To take advantage of faster LCVR modulators a new CCD camera ``Sensicam'' from M/s PCO, is procured and integrated into SVM during phase III. This will also help in making spectral scan faster. Further, chromospheric vector field measurement is planned using Na D1 line. The filters have been procured and calibration of LCVRs for this line is under progress.

\section{Conclusions}    
\label{sec:conclusions}
Comparison with HINODE-SOT Spectro-Polarimeter (SP) observations shows that reliable magnetic field parameters can be derived with SVM. The up-gradation to higher cadence will make the observations very valuable for studies of magnetic field evolution in relation to flares, line profile changes during flares, high frequency magnetic oscillations and waves in active regions and evolution of magnetic energy and helicity in regions with rapid flux emergence.


\begin{acknowledgement}
  We thank the conference organisers for a very good meeting and the
  editors for excellent instructions. We also thank the editor for giving beautiful Latex tips.
\end{acknowledgement}

\begin{small}


\begin{thebibliography}{5}
\bibitem[{{Beck} {et~al.}(2005){Beck}, {Schmidt}, {Kentischer}, \&
  {Elmore}}]{2005A&A...437.1159B}
{Beck}, C., {Schmidt}, W., {Kentischer}, T., {Elmore}, D. 2005, \aap, 437, 1159
\bibitem[{{Gosain} {et~al.}(2004){Gosain}, {Venkatakrishnan}, \&
  {Venugopalan}}]{2004ExA....18...31G}
{Gosain}, S., {Venkatakrishnan}, P., {Venugopalan}, K. 2004, Experimental
  Astronomy, 18, 31
\bibitem[{{Gosain} {et~al.}(2006){Gosain}, {Venkatakrishnan1}, \&
  {Venugopalan}}]{2006JApA...27..285G}
{Gosain}, S., {Venkatakrishnan}, P., {Venugopalan}, K. 2006, Journal of
  Astrophysics and Astronomy, 27, 285
\bibitem[{{Gosain} {et~al.}(2008){Gosain}, {Tiwari}, {Joshi}, \&
  {Venkatakrishnan}}]{2008JApA...29..107G}
{Gosain}, S., {Tiwari}, S., {Joshi}, J., {Venkatakrishnan}, P. 2008, Journal of
  Astrophysics and Astronomy, 29, 107
\bibitem[{{Socas-Navarro}(2001)}]{2001ASPC..236..487S}
{Socas-Navarro}, H. 2001, in Advanced Solar Polarimetry -- Theory, Observation,
  and Instrumentation, ed. M.~{Sigwarth}, Astronomical Society of the Pacific
  Conference Series, 236, 487
\end{thebibliography}

\end{small}

\end{document}